\documentclass[runningheads]{llncs}

\usepackage{graphicx}
\usepackage{xspace}
\usepackage[tagged, highstructure]{accessibility}

\usepackage[inline]{enumitem}

\usepackage{dirtytalk} 

\usepackage{tcolorbox} 
\newtcolorbox{mybox}[3][]
{
  colframe = #2!15,
  colback  = #2!2.5,
  coltitle = #2!20!black,  
  title    = {\textbf{#3}},
  #1,
}

\usepackage{xcolor}
\usepackage{colortbl}

\newcommand{\numParticipants}{nine\xspace}
\newcommand{\numCategories}{five\xspace}
\newcommand{\rqText}{What are instructors' expectations of a programming conversational agent intended to scaffold computational thinking?}

\newcommand{\ourparagraph}[1]{\vspace{4mm}\noindent\textbf{#1}}

\begin{document}

\title{Anticipating User Needs: Insights from Design Fiction on Conversational Agents for Computational Thinking}

\titlerunning{Anticipating User Needs: Insights from Design Fiction}

\author{Jacob Penney\inst{1}\orcidID{0009-0004-1629-5883}\and \\
João Felipe Pimentel\inst{2}\orcidID{0000-0001-6680-7470} \and \\
Igor Steinmacher\inst{1}\orcidID{0000-0002-0612-5790} \and \\
Marco A. Gerosa\inst{1}\orcidID{0000-0003-1399-7535}}

\authorrunning{J. Penney et al.}

\institute{Northern Arizona University\\
\email{\{jacob\char`_penney, igor.steinmacher, marco.gerosa\}@nau.edu} \\Universidade Federal Fluminense\\
\email{jpimentel@ic.uff.br}}

\maketitle

\vspace{-10pt}

%
%
\begin{abstract}

Computational thinking, and by extension, computer programming, is notoriously challenging to learn. Conversational agents and generative artificial intelligence (genAI) have the potential to facilitate this learning process by offering personalized guidance, interactive learning experiences, and code generation. However, current genAI-based chatbots focus on professional developers and may not adequately consider educational needs. Involving educators in conceiving educational tools is critical for ensuring usefulness and usability. We enlisted \numParticipants{} instructors to engage in design fiction sessions in which we elicited abilities such a conversational agent supported by genAI should display. Participants envisioned a conversational agent that guides students stepwise through exercises, tuning its method of guidance with an awareness of the educational background, skills and deficits, and learning preferences. The insights obtained in this paper can guide future implementations of tutoring conversational agents oriented toward teaching computational thinking and computer programming.


\end{abstract}

\section{Introduction} \label{sec:intro}

Knowledge of computer programming is key for modern society and for students~\cite{prensky2008programming}. Professionals from a diverse variety of industries need to write programs for tasks such as making financial predictions, office work, scientific research, creating entertainment, etc.~\cite{myers2006invited,lieberman2006end,ko2004six,burnett2009end,ghaoui2005encyclopedia}. End-user programmers---those who program but are not professional developers---have grown enough as a population to compel large tech firms to make major investments into technologies intended to ease development for them \cite{kuhail2021characterizing}. This increasing need among future workforce professionals to learn how to program is reflected in their educational needs today. It is not surprising that a large number of undergraduate programs include introductory programming in their curriculum. In fact, the literature shows plenty of evidence that STEM students perceive programming as a key skill in their careers~\cite{chilana2015perceptions,chilana2016understanding,sax2017examining,camp2015booming}. 

However, computer programming is difficult to learn~\cite{ko2003development,ko2004six,lewis1987can,pea1984cognitive,soloway2013studying}. Both undergraduate Computer Science majors and the growing population of non-majors who program struggle and show clear signs of poor performance, frustration, and lack of engagement~\cite{dawson2018designing,bosse2017programming,denny2011understanding}. Some institutions have reported dropout rates up to 50\%~\cite{kinnunen2006students}, and the estimated mean global pass rate for introductory Computer Science courses is around 68\%~\cite{outi2023attrition}. Substantial effort has been made to discover why learning how to program is chronically problematic. While there is no definitive consensus in the literature on what factors determine performance in introductory CS courses \cite{outi2023attrition}, many findings highlight student frustrations with programming language syntax, which is rigid and allows only a very restricted set of operations, as a reason for why learning is difficult~\cite{edwards2020syntax,denny2011understanding,kummerfeld2003neglected,lister2011computing,stefik2013empirical}. In interviews with students, Petersen et al.~\cite{petersen2016revisitingkinnunen} discovered a variety of contributing factors that prompt class withdrawal, such as the usage of ineffective study strategies; falling behind in the class and consequently receiving less support from in-class exercises and labs; and difficulty handling the relatively high level of detail in the material, which required developing problem-solving skills and was eased when they had access to \say{step-by-step instructions}. Student participants in other works also report difficulty adapting to the demands of CS1 courses, citing difficulty with \say{new teaching methods emphasizing independent thinking, critical thinking, and innovative learning}, as well as \say{less interaction with teachers and classmates}, \say{disconnection between the knowledge and real-life cases, and not receiving enough academic support with timely help and real-time feedback}~\cite{cao2023leveraging}.

New study strategies built on novel technologies, such as large language model-based (LLM-based) conversational agents -- that allow exercising computational thinking in a conversational way -- have the potential to overcome traditional strategies, allowing more students to succeed and grow in introductory programming courses. Such tools have achieved incredible performance even on complex assignments~\cite{nguyen2022empirical,li2022competition,prenner2021automatic,sobania2021choose,denny2023conversing}. GPT-3 has been used to create explanations of code snippets that are \say{significantly easier to understand and more accurate summaries of code} than what can be produced by first-year CS students \cite{leinonen2023comparing}, providing students on-demand access to explanations that explain code, freeing up instructor time. GPT-4 has even displayed rivaling human tutor performance in various programming education scenarios \cite{phung2023generative}. Not only are these tools capable, but findings show that users feel that they improve their outcomes. For example, programmers of various skill levels who used or were exposed to the features of a custom LLM-based conversational agent perceived that it could improve their productivity \cite{ross2023programmer}, and students feel \say{more motivated to learn, more engaged in the course, and more connected to their classmates}~\cite{cao2023leveraging}.

While evidence displays that LLM-based conversational agents can improve various outcomes for developers and students of different skill levels, little research exists on how to design such technology to improve the learning of computational thinking. Most existing solutions that allow for using a conversational way to program expose the artificial intelligence of the tool in a way not tailored to facilitate \textit{learning} to program. This is because they typically focus on professional productivity, giving the user the desired solution instead of using coaching and scaffolding, in the Cognitive Apprenticeship Model usage~\cite{collins1987cognitive}, to teach the user. Such tools could be used to refocus students' initial efforts on learning computational thinking \cite{luxton2016learning,becker2023programming} instead of focusing on implementation. The literature shows that implementation details can frustrate students because of the rigid set of operations allowed by programming language syntax~\cite{edwards2020syntax,denny2011understanding,kummerfeld2003neglected,lister2011computing,stefik2013empirical}. 

With this in mind, this work seeks to understand instructors' expectations of a conversational agent's capabilities in effectively facilitating the acquisition of computational thinking skills. The research question we answer here is:

\begin{mybox}{blue}{Research Question}
\rqText
\end{mybox}

%
To answer this question, we used the Design Fiction method.  
We analyzed the data collected from sessions with instructors by means of open and axial coding procedures. From this analysis, we learned that instructors expect a conversational agent intended to scaffold computational thinking to approach students with an awareness of their educational history and context and tailor its response to their questions accordingly. These responses should guide the student, maximizing skill set development. We hope these findings will inspire future research, design, and evaluation of conversational agents as well as serve as criteria for evaluating AI-based generative agents.
\vspace{-6pt}

\section{Related Work}
\vspace{-8pt}

After the recent proliferation of genAI tools, there have been escalating salvos of works exploring their use cases, quantifying and qualifying their abilities, and speculating about futures in which they have become fixtures. This work is situated among others which explore the perspectives of instructors about the experiences of students using conversational agents empowered by genAI to aid the learning process, the imagined and real benefits, and the potential pitfalls. Phung et al. \cite{phung2023generative} quantify the skills of existing LLM's, displaying not only where the state-of-the-art is at but also discussing where it is headed. Becker et al. \cite{becker2023programming} motivate educators in the booming genAI discussion, conveying an urgency to concertedly influence coming opportunities. Maher et al. \cite{MAHER2023ISC} propose and acknowledge many of the same benefits and concerns, respectively, that our participants imagined and introduce a methodology for analyzing AI tool impact via examining impact on student experience and abilities. Guo \cite{guo2023six} discusses ways that genAI can already be used for programming autodidacticism and considers ways that scientists and engineers can ply them for the educational needs of their specific fields.

The closest work to ours was performed by Lau et al. \cite{lau2023ban}. They ran interviews with instructors to understand how they plan to handle the use of genAI tools in their introductory programming classes. They found that the most common reason that instructors do not want to use AI-based tools in their classes is because they \say{felt it is still important to learn the fundamentals of programming, even if AI tools will be doing a lot of the coding in the future}. This opposition depends completely on the diegesis used, which asks the participant to imagine a tool that handles programming for the student with consistent perfect output, not paradigms that run counter to this \say{solution-oriented} functionality. Our work expands upon Lau et al. by taking some first steps towards answering open research questions they have posed, specifically \say{Scaffolding novice understanding} and \say{Tailoring AI coding tools for pedagogy}. We did this by approaching a population of instructors, as they did, with a diegesis oriented towards future conversational agents that do not have the limitations of current LLM-based ones.

%
%

\vspace{-8pt}
\section{Research Design}
\vspace{-4pt}
\subsection{Research Approach}
\vspace{-6pt}

To explore the design space of the conversational agent, we employed Design Fiction~\cite{sterling2009cover}. This method features \say{the deliberate use of diegetic prototypes to suspend disbelief about change}~\cite{sterling2009cover} and envision and explain plausible futures~\cite{blythe2014research,lupton2017design,Harmon2017,encinas2016solution,lindley2016peer,Muller2018,linehan2014alternate}. Human-Computer Interaction studies often use this method to probe, explore, and critique future technologies~\cite{lau2023ban,ringfort2022kiro,wessel2022bots,Muller2018,blythe2016co}. Some researchers use design fiction to anticipate issues~\cite{blythe2016co} while others emphasize values related to new technologies~\cite{liao,cheon} and anticipate users' needs~\cite{Cheon2017,encinas2016solution,noortman2019hawkeye}, which is the focus of this paper.

\vspace{-8pt}
\subsection{Method} \label{sec:method}
\vspace{-15pt}
\ourparagraph{The Fictional Narrative.}
We started the design fiction session by presenting a fictional narrative to the participants through a video~\cite{designfictionvideo2023}.
The video was intended to prompt participants to think about how a conversational agent could effectively scaffold computational thinking for students in introductory computer programming courses. In this scenario, a fictional student named Luna is learning Computer Science and, while she has access to an instructor during her class periods, she is left to find her own answers after class. The context of the video is a future in which present technological limitations have been overcome, and she has access to ecumenical AI tools that can help her with her studies. One such tool, Atlas, is a conversational agent that may interact conversationally and have access to and awareness of the student's context through integrations, such as coding environments, execution artifacts, exercises, and the progression of the course. The tool is presented as intended to help the student express her thoughts computationally. Having placed themselves in this narrative, we asked the participant to help us define how the conversational agent should behave so that it can help the student learn.

\vspace{-7pt}
\ourparagraph{Participant Recruitment.}
To participate in the design fiction sessions, we recruited instructors who have at least one year of professional experience teaching Computer Science or computational thinking. To access this population, we started by recruiting from the pool of instructors from our personal network and we used snowball sampling to help find new participants.

We screened potential participants by looking at their personal or university websites and invited those who had experience teaching introductory programming classes. During the invitation, we also sent our consent form that discussed relevant information about the study, including stipulations that our interviews would be recorded but that the recording and any resulting transcripts would be deleted after our work concluded.

In total, nine instructors agreed to participate in the study, as presented in Table~\ref{tab:demographics}.

\begin{table}[t]
\centering
\caption{Demographics of participants. \emph{Exp.} indicates the experience in years of the participant. \emph{Uni. Type} indicates the type of the university and the highest education level that it grants.\label{tab:demographics}}
\scriptsize
\setlength{\tabcolsep}{1mm}
\begin{tabular}{cccccl}
\hline
ID & \multicolumn{1}{c}{\begin{tabular}[c]{@{}c@{}}Exp.\end{tabular}} & Gender & Country & \begin{tabular}[c]{@{}c@{}}Uni. Type\end{tabular}  & \multicolumn{1}{l}{\begin{tabular}[l]{@{}l@{}}Introductory Classes\end{tabular}} \\
\hline
P1 & 14 & F & USA & Public-PhD & Algorithms, Data Structures, Programming \\
P2 & 20 & M & Brazil & Public-MSc & Algorithms, App Programming \\
P3 & 3 & M & USA & Public-PhD & Web Programming, Programming \\
P4 & 2 & M & Brazil & Public-MSc & Algorithms, Programming \\
P5 & 15 & M & Brazil & Public-PhD & Programming \\
P6 & 1.5 & M & Brazil & Public-BSc & Data Structures, Programming, Object Orientation \\
P7 & 10 & F & Brazil & Public-PhD & Programming \\
P8 & 15 & M & Brazil & Private-BSc & Algorithms, Data Structures, Web Programming \\
P9 & 3.5 & M & USA & Public-PhD & Algorithms, Data Structures, Programming \\
\hline
\end{tabular}
\end{table}

\vspace{-7pt}
\ourparagraph{Design Fiction Sessions.}
Qualified participants were invited to a meeting wherein one researcher introduced the goals of the work, displayed the two-minute prompt diegesis video, and interviewed the participants about how the conversational agent could help scaffold computational thinking and how they feel about such a tool. These interviews lasted for 30-60 minutes and were guided by these questions:
\begin{enumerate*}
    \item How can Atlas help Luna with computational thinking?
    \item How should Luna interact with Atlas?
    \item Where should Atlas be integrated to offer the best teaching experience?
    \item Are there any statistics that Atlas should collect from students and give to instructors?
    \item What benefits do you foresee with the use of Atlas?
    \item What drawbacks do you foresee with the use of Atlas?
    \item If you had two competing personal assistants, which criteria would you use to choose which one to use with your students?
    \item Do you have other suggestions about how Atlas should work? 
\end{enumerate*}

\begin{figure}[htbp]
    \centering
    \includegraphics[width=\linewidth]{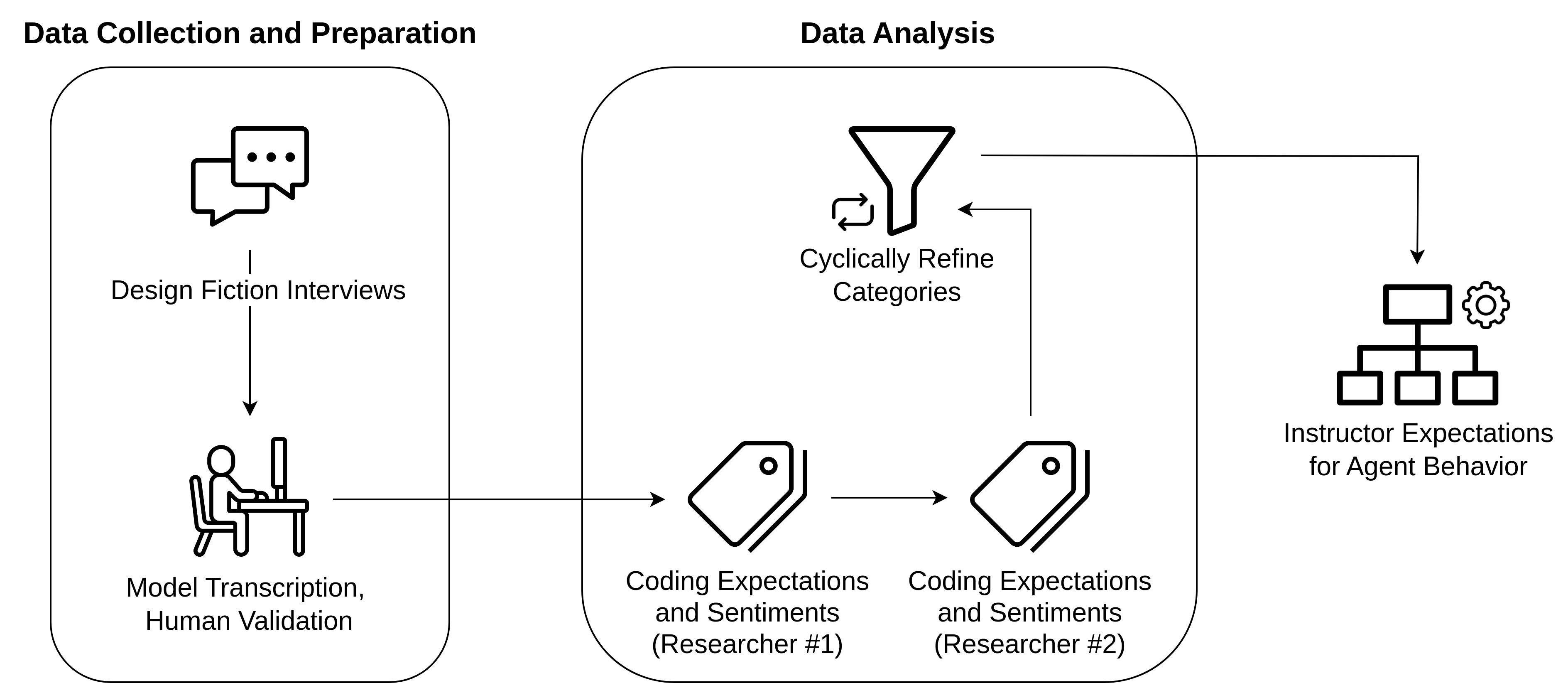}
    \caption{Our research method consists of five main steps, with the outcome of one step being the input to the next.}
    \alt{Basic flowchart of the paper's procedure. The first step of our research method was conducting design fiction interviews. Second, we used a speech recognition model to transcribe the interview text, after which a researcher listened to the interview and corrected errors in the transcription. Third, a researcher coded the corrected transcription. Fourth, a second researcher coded the same content. Fifth, one researcher cyclically refined the code categories. This process yielded our final output.}
    \label{fig:method}
\end{figure}

\ourparagraph{Analysis Method}
To analyze the narratives, as in similar studies~\cite{wessel2022bots,lau2023ban}, we applied open and axial coding procedures~\cite{Corbin2008Basics} through multiple rounds of analysis (see figure \ref{fig:method}). This analysis aimed to collect design insights and expectations that serve as guidelines for implementing such a conversational agent. After concluding an interview, the resulting recording was transcribed using a speech recognition tool~\cite{radford2023robust}, and a researcher reviewed the transcription with the recording to correct errors. Then, a researcher analyzed the text and coded participant quotes, after which a second researcher reviewed the first coding and applied coding a second time. After these two rounds of coding, we iteratively and cyclically reviewed and refined our categories, initially as a team and then conducted by the lead researcher. High-level categories of participant expectations began to emerge from our interview data and began to stabilize as our number of participants grew, indicating advancement towards data saturation, where these high-level categories nearly stopped evolving. We stopped interviewing participants since the final interviews did not bring significant insights to our results.

%
%
\vspace{-7pt}
\section{Results}
\vspace{-10pt}
In this section, we present our participants' expectations for an educational conversational agent and their sentiments about those expectations.

\vspace{-6pt}
\subsection{Expectations}
\begin{figure}[htbp]
    \centering
    \includegraphics[width=\linewidth]{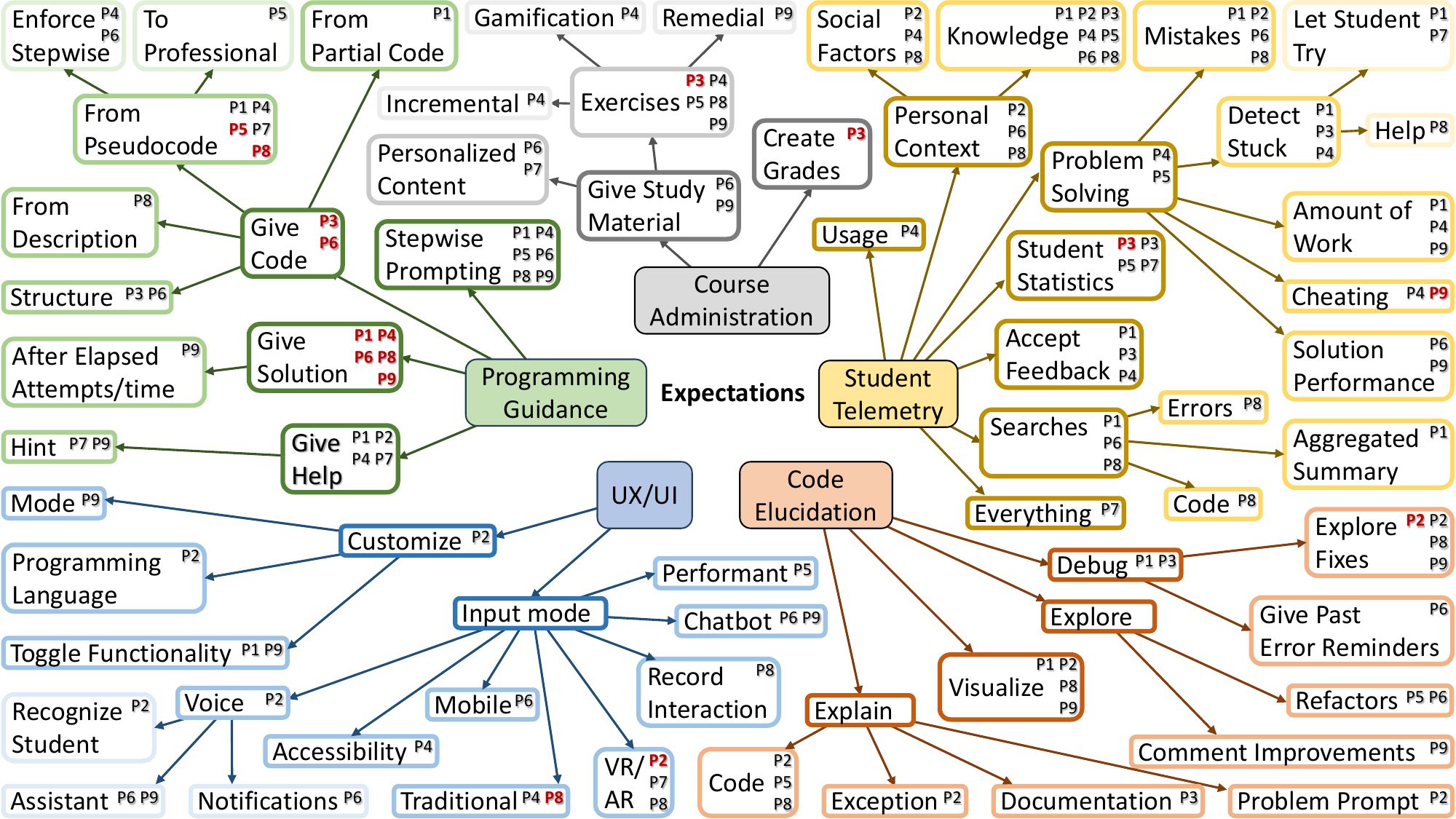}
    \caption{Categorization of expectations narrative data}
    \alt{Graph displaying hierarchical codes used to organize and understand the data.}
    \label{fig:expectations}
\end{figure}

\vspace{-6pt}
Participants described \numCategories categories of expectations that they felt would make the tool optimal for the end goal of scaffolding computational thinking for novice Computer Science students, displayed in Figure \ref{fig:expectations}: \textit{programming guidance}, \textit{code elucidation}, \textit{student telemetry}, \textit{course administration}, and \textit{UI/UX}. 

\ourparagraph{Programming Guidance.}
Instructors expect conversational agents to be able to scaffold computational thinking for students by guiding them stepwise through the development of algorithms in natural language instead of giving students solutions. The general idea was elaborated most lucidly by participant P1: \textit{\say{So, there are some standard questions that the teacher sometimes asks to say, `Look at what you want to do. What is the next step you would need to take in order to express yourself in the language you're using, in natural language, so to speak?'}}. While doing this, the agent should approach the student with an awareness of their educational background and knowledge and tailor their responses accordingly, such as explaining a concept differently to address the student's misunderstanding (P3). Support for giving the student solutions was overwhelmingly negative, and support for giving code was mixed and conditional, with four viewing it unfavorably (P3, P5, P6, P8) and three in support of translating pseudocode into code (P2, P4, P7). P4 felt that it was permissible because the intention is that students should focus on algorithmic thinking, not implementation. P3 felt it was permissible that the conversational agent give an algorithm in code that models the current problem, but which is not the solution and must be synthesized with their own ideas to successfully complete.

\ourparagraph{Code Elucidation.}
The conversational agent should be able to explain things in the environment that the student might be confused about, such as problem statements, code snippets, diagnostics, compilation and runtime errors, and underlying algorithms and concepts, such as memory. It should also be able to use diverse means to do so. For example, the most popular elucidation expectation that participants had was visualization, such as for algorithms (P1, P2, P9), memory (P8), debugging (P1), and code flow (P2). P2 emphasized \textit{\say{I understand that, from my experience, one of the best ways to make the student understand computational thinking and how it works is by drawing diagrams.}} Other examples of envisioned elucidation behaviors include retrieving documentation, highlighting passages relevant to the problem being addressed, and even explaining them if the student does not understand it (P3) and proactive and reactive debugging (P1, P2, P3, P8, P9) and refactoring (P5, P6), wherein the agent explains what can or should be done and why.

\ourparagraph{Student Telemetry.}
Participants acknowledged that many of the behaviors they expect from a conversational agent may require it to collect and use information about the student, and some desired for the agent to return statistics about the class for their consideration. Our participants thought of various pieces of information that would be useful to the agent and them. The most discussed was information about the student's formal and informal educational history (P2, P3), field of study or career (P4), doubts or struggles with course topics (P4, P5, P6), past usage of the tool (P2), and ability and disability (P3, P8). As discussed before, having information about the student could allow the agent to tailor responses, ideally to a high level. P3 explained that \textit{\say{there are many different ways to get to a solution, and if a student had that particular background, for instance, and Atlas was aware of that background, that might really help Atlas to maybe redirect the student to a different angle or a different approach to maybe better explore their lack of understanding on a topic}}. 

As per our participants, Atlas should display to the professor which topics students are struggling with, not only to help the current course but to create an awareness of specific struggles for the next iteration of the course. P6 thinks that \textit{\say{[the agent should collect statistics and create a] categorization of the main difficulties, so that... in a new class, the instructor can already have an idea of what they should address...}}. Participants also imagined that the conversational agent could observe the problem-solving behaviors of the student, such as detecting when they were stuck, how many and what kind of mistakes they made, the amount of effort and time the student put into work on the platform, the performance of their code solutions (maybe similar to how tools like Leetcode evaluate the speed of a solution), and whether or not the student cheated. 

To the contrary and looking towards the future of pedagogy, P3 felt that collecting statistics on student performance facilitates usage of outmoded methods of assessment: \say{the idea that students can navigate their own learning process at their own speed is entirely novel to... modern education. And I think that especially with tools like Atlas, we should really be stepping back from the concept of performance and especially like temporal performance achievements.}

\ourparagraph{Course Administration.}
Concerning course administration, participants were interested in the creation of study materials and exercises that were predicated on their students' past education, work they have done with the conversational agent, and deficiencies with the current topic. Some participants (P6, P8, P9) envisioned students being given remedial exercises based on scores on assessments taken in the agent. P4 envisioned assessments advancing incrementally, as they would in curriculum, and felt this pattern of escalation lends itself well to gamification, which would act as a means to maintain engagement. P3 opposed the conversational agent giving preprepared exercises because they cannot address the individual's particular misunderstandings or tailor their help to fit them. They also opposed the agent creating \say{final grades for a class... or even final grades for a set of concepts} because it may lack the flexibility and nuance that humans display when assessing how well a student has learned.

\ourparagraph{UI/UX.}
Expectations for how the user interface should behave, how the student should interact with the agent, or what the user experience should be like were varied and inconsistent. Participants imagine interacting with conversational agents via voice (P2, P6, P9), traditional input methods, virtual/augmented reality (P7, P8), and mobile (P6), with some detractors. P8 felt that popular utilities, like search engines, already respond via text, so the method of IO was not interesting. P2 felt that virtual reality doesn't offer much at the moment, but may in the future. The ability to customize the agent was interesting to a few participants (P1, P2, P9), specifically the ability to toggle functionality through atomic settings. Another suggestion was the use of \say{modes} dedicated to common usages, such as taking exams (P9). One participant, P4, mentioned accessibility a couple of times, referencing students with physical disabilities, such as deafness, and neuroatypicality, such as ADHD, and discussed that it would be meaningful for the system to have flexibility with IO methods to include such students. 

\subsection{Sentiments}

In addition to expectations, the participants discussed three categories of sentiments they had about such a tool in response to questions 5-7.

\vspace{-5pt}
\ourparagraph{Benefits.}
All-in-all, participants saw benefits of using a conversational agent oriented towards education for both students and instructors, but particularly for engaging, emotionally buttressing, and increasing understanding for students. Participants felt the tool could continuously assist students as they worked, potentially even in an \say{omnipresent} way (P8). Through this, they could experience decreases in feelings of isolation or fending for oneself (P7) and anxiety from not being able to finish their work right away and while a teacher is not around (P9). P3 explained that by providing a guide that will be present to help decompose and guide them through problems \say{you're giving students an opportunity to explore the problem without freaking out and... giving up}, which they feel \say{is one of the biggest barriers to success for computer science students.} The benefits to instructors that we found discussed most were that the agent could provide them with metrics about the class and help to free up their time by assisting with responding to students (P1), allowing them to focus on other areas of the class (P9), or even allowing them to teach more students (P9).

\vspace{-5pt}
\ourparagraph{Drawbacks.}
Reliance on the agent, teaching students incorrectly, and deteriorating relationships in the classroom are among the worries that participants had about the agent. If the tool engages in proactive intervention while the student is incorrectly implementing something, as an instructor might, it may condition students to wait for help (P2, P4). If the design is poor and students are capable of getting to solutions without receiving scaffolding, the student may not learn or may not be able to perform without the tool (P1, P4, P5). Of course, the instructor may have to watch for signs of misuse or cheating (P1, P2, P3). Then again, if the design works, participants worried that students may lose connection to their instructor or feel that they no longer need class time (P1, P6, P9). P3 noted that the tool may be difficult to administer, and it may make finding instructors more difficult because the institution will need those who can operate the tool.

\vspace{-5pt}
\ourparagraph{Priorities.}
In the scenario where participants were deliberating over which conversational agent to use, proven effectiveness with teaching students, credibility, and customization would sway their decision the most. Participants sometimes remarked on specific qualities that would promote learning the most, such as knowing when it is appropriate to intervene to offer assistance (P5, P7), teaching as opposed to simply giving answers and offering a comfortable interaction to the student (P6) and choosing the most effective means of teaching a concept, such as using visuals (P8). Emphasis on the tool's credibility betrayed that the participants who mentioned it did not fully immerse in the diegesis when it proposed that present technical limitations were not a concern, but that concern is significant and present. Instructors felt that customization would offer the ability for the tool to cater to needs more finely, such as by adjusting language (P4); the instructor having the ability to toggle functionality when needed, such as for specific assignments or tasks (P1, P9); and the student being able to toggle functionality as well, so they can get the experience they want (P9).

\begin{mybox}{blue}{Answer to the Research Question}
\textbf{\rqText} \\
{\color{blue!30}}
\noindent \\
The conversational agent is expected to act as a competent guide which would allow the student to move at their own pace through guided algorithm implementations and have access to thorough explanations in a variety of media, adapted to the individual student's background. Instructors would like the agent to collect statistics about student performance, usage, and background to make other abilities possible. Correct scaffolding behavior and credible guidance stemming from good data are preeminent and will sway educators to use a given conversational agent.
\end{mybox}

\section{Discussion} \label{sec:implications}
\vspace{-6pt}
Some of the behaviors that the instructors expect can already be meaningfully realized by existing LLMs and conversational agents built on top of them, such as ChatGPT. As introduced in Section~\ref{sec:intro}, LLMs have displayed that they can outperform first-year Computer Science students in creating explanations of basic function definitions~\cite{leinonen2023comparing}. More advanced models have displayed approaching the efficacy of human tutors in tasks such as \say{providing hints to a student to help resolve current issues} and \say{generating new tasks that exercise specific types of concepts/bugs}~\cite{phung2023generative}, abilities that our participants also expected. However, these recent findings do not address the more abstract ability to scaffold problem-solving skills or algorithm implementation via piecewise, tailored guidance without giving code or pseudocode, for one major example. Taming the output of LLMs for educational purposes is still an open problem, and developers implementing conversational agents for this purpose must consider how this can be accomplished.

Instead of allowing the student unrestricted access to the LLM's bank of information, developers should consider ways to regulate the LLM so that the response models actual problem-solving processes or provokes the student into critical thinking. Developers should also recognize that it is important to instructors that this response be tailored to the students' background and learning style. One popular method of tailoring LLMs is by prompt engineering, or \say{prompt crafting} \cite{bull2023eduprac}, on the natural language input. LLM outputs are known to be \say{sensitive} to inputs, producing different results for the same questions posed with different phrasing, and this quality can be exploited to strongly improve output appropriateness for a conversational agent focused on a specific domain \cite{denny2023conversing}. 

In the current landscape, recent scholarship notes that novel conversational agent implementations using GPT-4, such as Khanmigo, are advancing the state-of-the-art \cite{markel2023gpteach} of teaching using LLMs. Khanmigo poses questions to the student to prompt problem-solving behavior and can use redirection to refocus the student on the current question if they try to circumvent it. Functionalities such as this are possible through prompt engineering; indeed, LLM pipelines that rely on prompt engineering under the hood appear to be the current standard response to the question of taming LLM output, as we see with established conversational agent implementations such as Github Copilot\footnote{https://github.blog/2023-07-17-prompt-engineering-guide-generative-ai-llms/} and cutting-edge LLM-based tools from industry research, such as Microsoft Lida\footnote{https://microsoft.github.io/lida/} \cite{dibia2023lida}. Nevertheless, some expectations that our participants had cannot yet be realized. Dynamically creating visualizations tailored to address students' specific misunderstandings or situations is still hard with current LLMs; regardless, it is one of the most anticipated abilities among our participants. 


Finally, researchers interested in expanding upon our work can also investigate how certain effects can be accomplished. For example, participants were interested in the idea that assisting students may make them feel emotionally supported. While some felt that this was a natural consequence of having access to a powerful resource such as Atlas, it is not clear whether this hypothesis holds or what the role of other aspects would be, such as non-traditional interaction (e.g., voice and visual representations) and social characteristics (e.g., race, gender, culture) in the emotional support of students. For instance, researchers at the intersection of education and gender have established that gender plays a significant role in student engagement and that women STEM students are more inclined to seek help from women STEM instructors \cite{solanki2018looking}. Researchers looking to increase the comfort and effectiveness of LLM-based conversational agents should acknowledge and integrate work that examines the intersection of student and instructor identity with course engagement. A variety of other social characteristics of chatbots \cite{chaves2021should} can be investigated in this context. Future studies building upon this work can also incorporate the students' perspectives. Such research could present the same scenarios and pose the same questions to discern potential discrepancies in expectations and sentiments.
\vspace{-8pt}



%
%
\section{Limitations}
\vspace{-10pt}

In this work, we focused on the perspectives of instructors who teach or have recently taught introductory computer science. This method follows in the footsteps of other recent works \cite{lau2023ban}, but is a limitation because our results are skewed toward the experiences and interests of one population, only half of the classroom dynamic. Having no student voices leaves this work lacking insight on how students expect conversational agents to behave to help them develop computational thinking skills more effectively. New research focused on the student perspective may reveal entirely different sets of expectations and sentiments than those discussed by our current participants.

Even with focusing exclusively on instructors, our sample population is not diverse. We recruited instructors who work at five universities in Brazil and one in the United States. Most (7) were Brazilian cisgender men. we had only two instructors who were cisgender women, both Brazilian and one who taught in the United States. Consequently, our outcomes are biased along gender, culture, and region, for example because we have no trans or non-binary participants, and none from or teaching in Asia, Africa, or Europe. None explicitly expressed that they lived with disabilities which influenced their perspective, nor did they explicitly express that their racialization was a factor in their responses. The gender distribution of our participant cohort, in particular, arguably reflects the homogeneous gender distribution of the field it drew from, which has and still underrepresents cisgender women. Additional research which intentionally explores the experiences of more diverse populations may yield concerns not touched upon here. Of particular interest are the experiences of marginalized populations and discovering ways such tools can serve those who are least served by current pedagogy and educational institutions. Besides the identity of the instructors, other meaningful limitations may include that all but one instructor taught at public universities.

Much like related studies that use design fiction or similar methods \cite{lau2023ban}, our participants often discussed the future and hypothetical LLM-based conversational agents of the future in terms of existing conditions and similar extant technologies. While our interview questions and initial video prompt encouraged participants to envision a tool without the limitations faced by those that exist now, the sentiments they expressed may reflect the reality that they know. This includes how they conceived of UI/UX, such as imagining the tool using input methods that are currently common, such as voice; struggles LLMs face, such as the relevancy of data that the model was trained upon; and even common fears about LLM-based tools negatively impacting economic conditions for instructors. 
\vspace{-10pt}

\section{Conclusion}
\vspace{-8pt}
In this paper, we presented the expectations and sentiments of instructors involved with teaching Computer Science on the various functionalities that a LLM-based conversational agent could offer to best serve them in their work. Instructors imagined a tool that can scaffold computational using insights into the individual they are instructing, providing accessible tailored education outside of the classroom. This paper's findings lay the foundation for the implementation of novel solutions or the improvement of existing ones that can cater to the academic community, as well as introduce lines of further investigation. Future work on this topic include a design fiction study with student populations, Wizard of Oz experiments intended to classify student intentions and how they are expressed in dialogue, and design and prototype implementation of the conversational agent. 

\vspace{-10pt}
\section{Acknowledgments}
\vspace{-8pt}
This work was supported by the National Science Foundation grants 2236198, 2247929, and 2303042. We thank Alexander Gustav Siegel for assistance with our coding process and the instructors who shared their valuable experience participating in our research.

%
%
\bibliographystyle{splncs04}
\bibliography{cleanrefs}

\end{document}